\documentstyle[11pt,a4wide]{article}
\begin{document}
\begin{center}
{\Large\bf Explicit Construction of First Integrals with Quasi-monomial 
Terms from the Painlev\'{e} Series}\\
\vskip0.5cm
{\bf Christos Efthymiopoulos$^1$, Tassos Bountis$^2$, and 
Thanos Manos$^2$}\\
\vskip0.5cm
$^1${\em Research Center for Astronomy and Applied Mathematics,
Academy of Athens, \\
Soranou Efessiou 4, 115 27 Athens, Greece\\
\vskip0.3cm
$^2$Center for Research and Applications of Nonlinear Systems,\\
Department of Mathematics, University of Patras}\\
\vskip0.3cm
\end{center}
\vskip0.3cm \noindent {\bf Abstract} {\small The Painlev\'{e} and weak 
Painlev\'{e} conjectures have been used widely to identify new integrable 
nonlinear dynamical systems. For a system which passes the Painlev\'{e} 
test, the calculation of the integrals relies on a variety of methods 
which are independent from Painlev\'{e} analysis. The present paper 
proposes an explicit algorithm to build first integrals of a dynamical 
system, expressed as `quasi-polynomial' functions, from the information 
provided solely by the Painlev\'{e} - Laurent series solutions of a 
system of ODEs. Restrictions on the number and form of quasi-monomial 
terms appearing in a quasi-polynomial integral are obtained by an 
application of a theorem by Yoshida (1983). The integrals are obtained 
by a proper balancing of the coefficients in a quasi-polynomial function 
selected as initial ansatz for the integral, so that all dependence 
on powers of the time $\tau=t-t_0$ is eliminated. Both right and left 
Painlev\'{e} series are useful in the method. Alternatively, the method 
can be used to show the non-existence of a quasi-polynomial first integral. 
Examples from specific dynamical systems are given.}

\section{Introduction}

The present paper deals with autonomous dynamical systems described by 
ordinary differential equations of the form 
\begin{equation}\label{ode1}
\dot{x_i}=F_i(x_1,x_2,...,x_n),~~~~~~i=1,2,...,n.
\end{equation}
where the functions $F_i$ are of the form
\begin{equation}\label{ode2}
F_i = \sum_{j=1}^{N}a_{ij}x^{Q_{ij}}
\end{equation}
with 
$x\equiv(x_1,x_2\ldots x_n)$, $Q_{ij}\equiv (q_{ij1},q_{ij2},\ldots,
q_{ijn})$, $x^{Q_{ij}}\equiv \prod_{k=1}^n x_k^{q_{ijk}}$, and the 
exponents $q_{ijk}$ are assumed to be rational numbers, i.e., the r.h.s. 
of Eq.(\ref{ode2}) is a sum of quasimonomials (Goriely 1992). 

A question of particular interest concerns the existence of {\it first 
integrals} of the system of ordinary differential equations (\ref{ode1}). 
A first integral is a function $I(x_i)$ satisfying the equality $dI/dt = 
\nabla_xI\cdot\dot{x}(t)=0$, where $x(t)\equiv (x_1(t),\ldots,x_n(t)$ 
is any possible solution of (\ref{ode1}). First integrals are important 
because they allow one to constrain the orbits on manifolds of dimensionality 
lower than $n$. In particular, a system (\ref{ode1}) is called completely 
integrable if it admits $n-1$ independent and single-valued first 
integrals. Integrable systems exhibit regular dynamics, while the lack 
of a sufficient number of first integrals very often results in complex, 
chaotic dynamics. 

The question of the existence of an algorithmic method which can 
determine all the first integrals of (\ref{ode1}) is an important open 
problem in the theory of ordinary differential equations. Most relevant 
to this question are the methods of a) direct search or method of 
undetermined coefficients (e.g. Hietarinta 1983, 1987) b) normal forms 
and formal integrals (see Arnold 1985, Haller 1999 and Gorielly (2001) 
for a review), and c) Lie group symmetry methods (e.g. Lakshmanan and 
Senthil Velan (1992a,b), Marcelli and Nucci (2003)).

Even more difficult is the question of an algorithmic method probing 
the integrability of Eqs.(\ref{ode1}). The most relevant method here 
is {\it singularity analysis}. According to the `Painlev\'{e} 
conjecture' (Ablowitz, Ramani and Segur 1980), a system possessing 
the Painlev\'{e} property should be integrable. The Painlev\'{e} 
property means that any global solution of (\ref{ode1}) in the complex 
time plane should be free of movable critical points other than poles. 
According to the `weak-Painlev\'{e} conjecture (Ramani et al. 1982, 
Grammaticos et al. 1984, Abenda et al. 2001), however, certain types of 
movable branch points are compatible with integrability. Algorithms providing 
necessary conditions for a system to be Painlev\'{e} are a) the classical 
ARS test (Ablowitz et al. 1980) b) the perturbative Painlev\'{e} - Fuchs 
test (Fordy and Pickering 1991) which examines the role of negative 
resonances, and c) the generalized Painlev\'{e} test of Goriely (1992) 
which introduces coordinate transformations clarifying the nature of the 
singularity structure in the complex time domain. On the other hand, there 
is no algorithm, to the present, determining sufficient conditions for a 
system to be Painlev\'{e}. The most important obstacle is the search for 
essential singularities, which are not detectable by any of the above 
Painlev\'{e} tests.

The present paper explores the following question: Is it possible to 
recover the first integrals of a system by the information provided 
solely by singularity analysis of its differential equations? 
Our answer is partially affirmative. We cannot circumvent the difficulty 
concerning the choice of initial ansatz for the functional form of the 
integral. The most natural choice for quasi-polynomial equations 
(\ref{ode2}) is to consider also a quasi-polynomial ansatz for the 
integral,  with undetermined quasi-monomial coefficients. 

This freedom in the initial ansatz notwithstanding, we show in the 
present paper that singularity analysis gives indeed the remaining 
information needed to recover the integrals. 

First, as proposed by Roekaerts and Schwarz (1987), a theorem by 
Yoshida (1983) on the relation between Kowalevski exponents and 
weights of weighted-homogeneous integrals, can be used to impose 
restrictions on the degree of the quasi-monomial terms in the 
integral by analysing the resonances found by the Painlev\'{e} method. 
Now, Yoshida's theorem for weighted-homogeneous integrals is applicable 
on two conditions: If $b_i\tau^{-\lambda_i}$ is a balance of the system 
($\tau=t-t_0$ is the time around the singularity $t_0$), and $I$ is 
a weighted homogeneous integral, the theorem holds if 
a) $\nabla I(b_i)$ is finite, and b) $\nabla I(b_i)\neq 0$. These 
conditions shall be refered to as `Yoshida's conditions'. The latter 
impose a severe restriction in the search for first integrals, because 
one cannot specify in advance whether these conditions are satisfied until 
the integrals are determined. In conclusion, the theorem of Yoshida has 
only indicative power as regards restrictions on the degree of 
quasimonomial terms in the integral. On the other hand, an analysis 
of the integrable Hamiltonian systems of two degrees of freedom with a 
polynomial potential given by Hiterinta (1983) shows that  they all 
satisfy Yoshida's condition when the balance is taken equal to the 
`principal' balance, in which all the $b_i$ are different from zero. 
Yet, the extent of applicability of this result to other types of systems 
is unknown. In fact, we were able to find also a counterexample 
concerning a Hamiltonian proposed by Holt (1982). 

Assuming a quasi-polynomial functional form of the integral, say 
$\Phi(x;c_1,c_2,...,c_M)$,as above, with undetermined quasi-monomial 
coeficients $c_i,i=1,\ldots M$, the question now is whether it is possible 
to determine the coefficients $c_i$ by the series derived via singularity 
analysis. The answer to this question is affirmative. Namely, by the usual 
Painlev\'{e} tests, the Painlev\'{e}-type series solutions around movable 
singularities are first identified 
\begin{equation}\label{pain1}
x_i(\tau) = {1\over \tau^{\lambda_i}}
\sum_{m=0}^{\infty}b_m\tau^m,~~~i=1,\ldots n
\end{equation}
Then, the series (\ref{pain1}) {\it is substituted} into the 
quasi-polynomial function $\Phi(x;c_1,c_2,...,c_M)$. The resulting 
expression is a Puisseux series, i.e., a series in rational powers of 
$\tau$
\begin{equation}\label{powtau}
\Phi(x_1(\tau),\ldots,x_n(\tau);c_1,c_2,...,c_M) = 
\tau^{q/p}\sum_{m=0}^{\infty} d_m(b_i;c_i)\tau^{m/p}
\end{equation}
with $q,p,m$ integers. The coefficients $d_m(b_i;c_i)$ are 
nonlinear functions of the coefficients $b_i$ (determined by (\ref{pain1}), 
i.e., by singularity analysis), and linear functions of the undetermined 
coefficients $c_i$. But the function $\Phi$ is an integral of the system 
if it is constant along all the solutions of the system, including 
(\ref{pain1}). As a result, the functions $d_m$ satisfy the set of linear 
equations
\begin{equation}\label{dm}
d_m(b_i;c_i) = 0,~~~~i=0,1,\ldots,~~~i \neq -q
\end{equation}
This is an infinite number of homogeneous linear equations with a 
finite number of unknowns (the coefficients $c_i$). If $M$ is the number 
of unknown coefficients, Eq.(\ref{dm}) can be written as
\begin{equation}\label{dm2}
A(b_i)\cdot C = 0
\end{equation}
where $C=(c_1,\ldots,c_M)^T$ and $A(b_i)$ is a matrix with $M$ columns 
and an infinite number of lines. The entries of $A$ depend only on the 
coefficients $b_i$ which were previously determined by singularity 
analysis. In this representation, the first integrals are functions 
with quasi-monomial coefficients given by the basis vectors of $ker(A)$ 
(or linear combinations of them). In computer algebraic implementations 
of the method, we work on a sub-matrix $A_f$ defined by a finite number 
of lines, which is equal or larger than $M$. A basis for the subspace 
$ker(A_f)$ is determined by the singular value decomposition algorithm. 
Then, it is checked with direct differentiation that the resulting 
expression is an integral. This completes the determination of all 
quasi-polynomial first integrals for the given system. 

This method was implemented in a number of examples presented below. 
Following some preliminary notions exposed in section 2, the results are 
presented in section 3, along with various details and implications in 
the implementation of the algorithm. Section 4 summarizes the main 
conclusions of the present study.

\section{Preliminary notions}

Following Yoshida (1983), a system of the form (\ref{ode1}) is called 
scale-invariant if the equations remain invariant under the scale
transformation $x_i\rightarrow a^{\lambda_i}x_i$, $t\rightarrow 
a^{-1}t$ for some $\lambda_i$. Then, the system (\ref{ode1}) 
has  exact special solutions of the form
\begin{equation}\label{balance}
x_i = {b_i \over \tau^{\lambda_i}}
\end{equation}
where $b_i,i=1,\ldots n$ is any of the sets of roots of the system of 
algebraic equations 
$$
F_i(b_1,b_2,...,b_n)+\lambda_ib_i=0
$$ 
and the time $\tau=t-t_0$ is considered around any movable singularity 
$t_0$ in the complex time plane. Any solution of the form (\ref{balance}) 
is called a `balance'. If the system (\ref{ode1}) is scale-invariant and 
the functions $F_i$ are of the form (\ref{ode2}), the exponents $\lambda_i$ 
associated with any of the balances are rational numbers. It should 
be stressed that the definition above does not require that all the 
$b_i$ be non-zero. Balances of the form $x_i\sim 0/\tau^{\lambda_i}$, 
for some $i$, are also considered. The latter remark is essential in 
order to avoid the confusion which is sometimes made between `balances' 
and `dominant terms' in the ARS test (see e.g. the discussion between 
Steeb et al. (1987) and Ramani et al. (1988)), and the associated 
difference between `Kowalevski exponents' and `resonances'. In the 
standard ARS algorithm (Ablowitz et al. 1980) the solutions (\ref{balance}) 
arise by the definition of the dominant behaviors, i.e., which is the 
first step in the implementation of the algorithm.

The second step in the ARS algorithm is to look for series solutions 
that we call 'Painlev\'{e} series'. These are expansions 
of the form (\ref{pain1}) starting with dominant terms of the form 
(\ref{balance}). They are Laurent (Taylor) series when the $\lambda_i$'s 
are integers (positive integers), otherwise they can be series in 
rational powers of $\tau$, which are called `Puisseux series'. To build 
up the series, one first specifies the resonances, i.e. the values of 
$r$ for which the coefficients of the terms $\tau^{r-\lambda_i}$ in 
the series are arbitrary. In the case when the coefficients of the 
balance $b_i$ are all non-zero, the resonances are equal to the 
eigenvalues of the  Kowalevski matrix 
$K_{ij}=(\partial F_i/\partial x_j + \delta_{ij}\lambda_i)|_{x_i=b_i}$ 
which are called `Kowalevski exponents'. If, however, some of the 
$b_i$s are equal to zero, then the resonances are not equal one to 
one to the Kowalevski exponents, but some resonances differ from the 
corresponding Kowalevski exponents by a quantity equal to the difference 
between the exponent $\lambda_i$ in the balance and the exponent of the 
first non-zero dominant term in the Laurent-Puisseux series of $x_i(\tau)$, 
as specified in the first step of the ARS algorithm (Ramani et al. 1988).

At this point, we are {\it not} interested in whether the system passes 
the generalized Painlev\'{e} (or weak Painlev\'{e}) test. This means 
that we do {\it not} require that {\it all} the solutions of the 
system (\ref{ode1}) can be written locally (around a movable 
singularity) in the form (\ref{pain1}), or that there is at least 
one solution of the form (\ref{pain1}) which contains $n$ arbitrary 
constants (including $t_0$). On the other hand, we {\it do} check 
the compatibility conditions to ensure that no logarithms enter in 
the series. As regards positive resonances, compatibility is fulfilled 
automatically for scale-invariant systems. In summary, 
we are interested only that the system have special solutions of 
the form (\ref{pain1}), but no other claim on it being Painlev\'{e} 
or not is required. Thus, the results are valid also for partially 
integrable systems, i.e., systems with a number of first integrals 
smaller than $n$. 

Let us now assume that (\ref{ode1}) possesses a weighted - homogeneous 
first integral $\Phi$ of  weight $M$, i.e. an integral function  
$\Phi$ which satisfies the relation:
\begin{equation}\label{weighted}
\Phi(a^{\lambda_1}x_1,a^{\lambda_2}x_2,...,a^{\lambda_n}x_n)=
a^M\Phi(x_1,x_2,...,x_n)
\end{equation}
for some $M$. Then we have the following\\
\\
\noindent{\bf Theorem 1} (Yoshida 1983): {\it If, for a particular balance 
(\ref{balance}) the following conditions hold: a)$\nabla\Phi(b_i)$ is finite, 
and b) $\nabla \Phi(b_i)\neq 0$, then $M$ is equal to one of the 
Kowalevski exponents associated with that balance}.\\
\\
There has been a number of theorems in the literature linking Kowalevski 
exponents with the weights of homogeneous or quasi-homogeneous integrals 
of a system. Some characteristic papers on this subject are Llibre and Zhang 
(2002), Tsygvinsev (2001), Goriely (1996) and Furta (1996). However, the 
application of Yoshida's original theorem in the search for 
weighted-homogeneous integrals seems to be the most practical. 
Furthermore, Yoshida's theorem can be reformulated in an interesting way: 
consider the Painlen\'{e} series starting with one of the balances 
(\ref{balance}). It follows that the series can be written as:
\begin{equation}\label{xiseries}
x_i=x_{iE}+R_i={b_i\over
\tau^{\lambda_i}}+...+A_i\tau^{r-\lambda_i}+R_i
\end{equation}
where $r$ is the maximum Kowalevski exponent associated with this 
balance and $A_i$ is the corresponding arbitrary coefficient entering 
in the Painlev\'{e} series (\ref{xiseries}) for the variable $x_i$. The 
sum $x_{iE}={b_i\over \tau^{\lambda_i}}+...+A_i\tau^{r-\lambda_i}$ 
will be called {\it essential part} of the series and the remaining 
part $R_i$ {\it remainder} of the series. The remainder $R_i$ starts 
with terms of degree $r-\lambda_i+1/p$ where $p$ is the denominator of 
$\lambda_i$ written as a rational $\lambda_i=q/p$ with $q,p$ coprime 
integers. A quasi-polynomial integral $\Phi$ has the form 
\begin{equation}
\Phi=\sum c_{k1,k2,...,kn}\prod_{i=1}^n x_i^{k_i}
\end{equation}
where the exponents $k_i$ are rational numbers. If the integral $\Phi$ 
is weighted-homogeneous of weight $M$, we have 
\begin{equation}
\sum_{i=1}^nk_i\lambda_i=M
\end{equation}
due to (\ref{weighted}). Taking into account the fact that the remainder 
$R_i$ starts with terms of degree $O(\tau^{r-\lambda_i+1/p})$, it 
follows that the contribution of $R_i$ in $x_i^{k_i}$ is in terms 
of degree $O(\tau^{-\lambda_ik_i+r+1/p})$ or higher. This means that 
if the series (\ref{xiseries}) is substituted in the integral 
$\Phi(x_i)$, the contribution of $R_i$ in $\Phi(x_i)$ is of 
degree $O(\tau^{-\sum\lambda_ik_i + r +1/p}) = O(\tau^{-M+r+1/p})$ or
higher. But $r\geq M$. Thus, the remainder $R_i$ contributes only
to terms of {\em positive} degree in $\tau$. On the other hand, 
since $\Phi$ is an integral, the time $\tau$ as a denominator must be 
eliminated in $\Phi(x_i)$. But since there are no negative powers of 
$\tau$ generated in $\Phi$ by $R_i$, it follows that all the negative 
powers of $\tau$ are already eliminated by substituting the expression 
$x_{iE}$ alone into $\Phi$. Hence, we have the following\\
\\
\noindent {\bf Proposition 2:} {\it If the conditions of Yoshida's theorem 
hold for a weighted-homogeneous integral of (\ref{ode1}) and a particular 
balance (\ref{balance}), then the expression $\Phi(x_{iE})$, where $x_{iE}$ 
are the essential parts of Painlev\'{e} series $x_i(\tau)$ initiated 
with the same balance, does not contain singular terms in $\tau$.}\\
\\
Consider next the case when the functions $F_i$ in (\ref{ode1}) 
are not homogeneous. By the restrictions imposed by (\ref{ode2}), it 
follows that the functions $F_i$ can be decomposed in sums of the 
form 
\begin{equation}\label{poly}
F_i=F_i^{(m_{i0})}+F_i^{(m_{i0}+1/p)}+...+F_i^{(m_{i0}+q/p)}
\end{equation}
where the functions $F_i^{(j)}$ are homogeneous of degree $j$, with $j,m_{i0}$ 
rational, $p,q$ integer, and $p$ is the denominator in the simplest fraction 
giving $m_{i0}$. In this case, if the system (\ref{ode1}) has 
Painlev\'{e} type solutions, the dominant behaviors and associated 
resonances of these solutions are  determined by the homogeneous term 
of the highest degree $F_i^{(m_{i0}+q/p)}$. On the other hand, the 
functions $F_i^{(j)}$, $j<m_{i0}+q/p$ must have a special form to ensure 
that compatibility conditions are fulfilled and the series solution 
is of the Painlev\'{e} type. Finally, as regards potential first integrals, 
the assumption that they consist of a sum of quasi-monomial terms implies 
that they can also be written as sums of the form (\ref{poly}). The 
selection of terms in the quasi-polynomial integral can be determined by 
Yoshida's theorem (or proposition 2) implemented in the scale-invariant 
systems
\begin{equation}\label{low}
\dot{x}_i=F_i^{(m_{i0})}(x_i)
\end{equation}
and 
\begin{equation}\label{high}
\dot{x}_i=F_i^{(m_{i0}+q/p)}(x_i)
\end{equation}
respectively (Nakagawa 2002).

\section{Explicit construction of integrals with quasi-monomial terms}

\subsection{An elementary example}

Consider the two-dimensional nonlinear system
\begin{equation}\label{sys1}
\dot{x}_1=x_2,~~~~~~\dot{x}_2=-x_1 -3x_1^2
\end{equation}
The only first integral of this system
\begin{equation}\label{int1}
\Phi=x_1^2+x_2^2  + 2x_1^3
\end{equation}
can be recovered by elementary means. However, we will use this example 
to illustrate the steps used by the present method. The corresponding 
homogeneous system containing the terms of maximum degree 
\begin{equation}\label{sys1h}
\dot{x}_1=x_2,~~~~~~\dot{x}_2=-3x_1^2
\end{equation}
has the unique balance $x_1=-2/\tau^2$, $x_2=4/\tau^3$, with Kowalevski 
exponents (equal to resonances) $-1$ and $6$. Combatibility conditions 
are fulfilled for the system (\ref{sys1}), which admits the Laurent 
series solution
\begin{eqnarray}\label{laur1}
x_1(\tau) &= &{-2\over\tau^2} - {1\over 6} - {1\over 120}\tau^2 +
            a_4\tau^4 + O(\tau^6) \nonumber \\
x_2(\tau) &= &{4\over\tau^3} - {1\over 60}\tau + 4a_4\tau^3 + O(\tau^5)
\end{eqnarray}
where $a_4$ is an arbitrary parameter.

Following Yoshida's theorem, we shall look for an integral of the system 
(\ref{sys1}) by requesting that this integral be a sum of weighted-
homogeneous functions of weight not higher than $M=6$. Since the equations 
(\ref{sys1}) are polynomial, the integral will also be assumed polynomial. 
According to the definition of the weighted-homogeneous functions 
(\ref{weighted}), the undetermined integral contains terms of the 
form $x_1^{q_1}x_2^{q_2}$ the exponents of which are restricted by the 
relation $2q_1+3q_2\leq 6$. This leaves only six possibilities, namely 
$(q_1=1,q_2=0)$, $(q_1=0,q_2=1)$, $(q_1=2,q_2=0)$, $(q_1=1,q_2=1)$, 
$(q_1=0,q_2=2)$, $(q_1=3,q_2=0)$. Thus the integral is assumed to 
have the form 
\begin{equation}\label{int1n}
\Phi = c_{10}x_1 +c_{01}x_2 + c_{20}x_1^2 +
c_{11}x_1x_2 + c_{02}x_2^2 + c_{30}x_1^3
\end{equation}
Up to now, the steps are exactly as proposed by Roekaerts 
and Schwarz (1987). At this point, however, we do not proceed 
by the `direct method'; instead, the series (\ref{laur1}) is 
substituted into (\ref{int1n}). Then, terms of equal power in $\tau$ 
are separated and their respective coefficients are set equal to 
zero. We must determine at least as many equations as the 
number of unknown coefficients $c_{ij}$, i.e., six equations. 
These are:
$$
Order~~O(1/\tau^6):~~~~~~~~~~~~~~~~~~~~~~~~~~~~16c_{02}-8c_{30}=0
$$
$$
Order~~O(1/\tau^5):~~~~~~~~~~~~~~~~~~~~~~~~~~~~~~~~~~~~-8c_{11}=0
$$
$$
Order~~O(1/\tau^4):~~~~~~~~~~~~~~~~~~~~~~~~~~~~~~4c_{20}-2c_{30}=0
$$
$$
Order~~O(1/\tau^3):~~~~~~~~~~~~~~~~~~~~~~~~~~~~~~4c_{01}-{2\over 3}c_{11}=0
$$
$$
Order~~O(1/\tau^2):~~~~-2c_{10}+{2\over 3}c_{20}-{2\over 15}c_{02}-{4\over 15}c_{30}=0
$$
$$
Order~~O(1/\tau):~~~~~~~~~~~~~~~~~~~~~~~~~~~~~~~~~~~~~~~~~~~~~0=0
$$

These equations can be written in matrix form:
\begin{equation}
\left(
\begin{array}{cccccc}
0 &0 &0 &0 &16 &-8 \\
0 &0 &0 &-8 &0 &0 \\
0 &0 &4 &0 &0 &-2 \\
0 &4 &0 &-2/3 &0 &0 \\
-2 &0 &2/3 &0 &-2/15 &-4/15\\
0 &0 &0 &0 &0 &0 \\
\end{array}
\right) \left(
\begin{array}{c}
c_{10}\\
c_{01}\\
c_{20}\\
c_{11}\\
c_{02}\\
c_{30}\\
\end{array}
\right)
=\left(
\begin{array}{c}
0\\
0\\
0\\
0\\
0\\
0\\
\end{array}
\right)
\end{equation}
or simply
\begin{equation}
A_f\cdot C=0
\end{equation}
where $C$ is a six-dimensional vector and $A_f$ is a $6\times 6$ 
matrix with constant entries. 

The singular value decomposition of $A_f$ yields a one-dimensional 
null space:
\begin{equation}\label{ker1}
ker(A_f)=\lambda(0,0,1,0,1,2,0)
\end{equation}
The basis vector $(0,0,1,0,1,2)$ corresponds to the first integral 
$\Phi=x_1^2+x_2^2+2x_1^3$.

A few remarks are here in order: 

a) the last line of $A_f$ has only zero entries, since it corresponds 
to the identity $0=0$ for the $O(1/\tau)$ terms. This is not a problem, 
because lines with zero elements are allowed by the singular value 
decomposition algorithm which determines the subspace $ker(A_f)$. 

b) The entries of $A_f$ are constant numbers which depend only on 
the coefficients of the Laurent series (\ref{laur1}). This is the 
crucial remark; it implies that the information on the first integral 
is contained in the Painlev\'{e} series built by singularity analysis.   

c) The arbitrary parameter $a_4$ in (\ref{laur1}) does not appear 
in $A_f$. This phenomenon is not generic. In general, all the arbitrary 
parameters of the Painlev\'{e} series appear in $A_f$. In the computer 
implementation of the algorithm, we proceed by giving fixed values to 
the arbitrary parameters. Although the choice of values affects the 
convergence of the Painlev\'{e} series, it does not influence the 
present algorithm which is based only in the formal properties of the 
series. 
\subsection{Further examples}

Of particular interest in nonlinear dynamics are autonomous 
Hamiltonian systems of two degrees of freedom of the form
\begin{equation}\label{ham1}
H\equiv {1\over 2}(p_x^2+p_y^2)+V(x,y)
\end{equation}
where $V(x,y)$ is of the form (\ref{ode2}). The easiest examples 
are systems with a polynomial potential (e.g. Hietarinta 1983, 
1987). For example:
\begin{equation}\label{hambnt}
H\equiv {1\over 2}(p_x^2+p_y^2+x^2+y^2)-x^2y-2y^3
\end{equation}
This system passes the Painlev\'{e} test and it is integrable 
(Bountis et al., 1982). Keeping only the highest order terms of the 
potential ($-x^2y-2y^3$) yields the principal balance 
$x=6i/\tau^2, y=3/\tau^2$ with resonances $r=-3,-1,6,8$, which are 
equal to the corresponding Kowalevski exponents. The Painlev\'{e} 
series generated by the Hamiltonian (\ref{hambnt}) and the above 
principal balance satisfies the compatibility conditions of the ARS test. 
Assuming now a polynomial first integral $\Phi$ of (\ref{hambnt}), only 
the monomial terms of weight less or equal to 8 will be included to it. 
Since the leading terms of the momenta are $p_x\sim p_y\sim O(1/\tau^3)$, 
the selected monomial terms are:
$$
x^4,~x^3y,~x^2y^2,~xy^3,~y^4,~
p_x^2x,~p_xp_y x,~p_y^2 x,~p_x^2 y,~p_xp_y y,~p_y^2y~~~\mbox{(weight 8)}
$$
$$
p_xx^2,~p_xxy,~p_xy^2,~p_yx^2,~p_yxy,~p_yy^2,~~~~\mbox{(weight 7)}
$$
$$
x^3,~x^2y,~xy^2,~y^3,~p_x^2,~p_xp_y,~p_y^2,x^3,~x^2y,~xy^2~,y^3~~~\mbox{(weight 6)}
$$
$$
p_xx,~p_xy,~p_yx,~p_yy,~~~\mbox{(weight 5)}
$$
$$
x^2,~xy,~y^2,~p_x,~p_y,~x,~y~~~\mbox{(weights 4,3,2)}
$$
It should be stressed that there are several other restrictions that 
reduce the number of eligible terms. For example, the integral $\Phi$ 
is either even or odd in the momenta (Nakagawa and Yoshida 2001). 
Furthermore, the linear terms can be omitted by appropriate 
transformations. However, in the practical implementation of the 
algorithm these restrictions only introduce a complication, because, 
if an integral $\Phi$ exists, the singular value decomposition algorithm 
selects the basis for the corresponding null space of the matrix $A_f$ 
without needing any extra information on restrictions which are specific 
for the system under study, e.g. hamiltonian or other. 

Following the selection of monomial terms, the algorithm proceeds in 
building the homogeneous system (\ref{dm2}) as well as the finite 
restriction $A_f$ of the matrix $A$. In this case, the dimension 
of $A_f$ should be $M\times 38$, with $M\geq 38$, since there are 
38 unknown coefficients of the above monomial terms. At this point, 
it does not matter which balance and Laurent-generated series 
are used to build the matrix $A_f$. In the above example, the principal 
balance leads to a 'special solution', since there are only two 
positive resonances ($r=6,8$) meaning that there are three arbitrary 
parameters in total entering in the series. On the other hand, 
the general Laurent series solution with four arbitrary constants 
is given by a different balance, namely $x=0/\tau^2,y=1/\tau^2$ 
so that $x$ starts with dominant terms of $O(1/\tau)$, i.e.
\begin{eqnarray}\label{laubnt}
x(\tau) &= &{A\over\tau} + a_1\tau + B\tau^2 + a_3\tau^3 + a_4\tau^4
            a_5\tau^5 + O(\tau^6) \nonumber \\
y(\tau) &= &{1\over\tau^2} +b_0 + b_2\tau^2 + b_3\tau^3 
+ C\tau^4 + O(\tau^5) \\
p_x(\tau) &= &-{A\over\tau^2} + a_1 + 2B\tau + 3a_3\tau^2 + 4a_4\tau^3
            5a_5\tau^4 + O(\tau^5) \nonumber \\
y(\tau) &= &-{2\over\tau^3} + 2b_2\tau + 3b_3\tau^2 
+ 4C\tau^3 + O(\tau^4) \nonumber
\end{eqnarray}
where $A,B,C$ together with $t_0=t-\tau$ are arbitrary, and
$b_0 = (1-A^2)/12$, $a_1=A(1-2b_0)/2$, $b_2=(b_0-6b_0^2-2Aa_1)/10$,
$a_3=(2b_0a_1-a_1+2Ab_2)/4$, $b_3=-AB/3$, $a_4=(2Ab_3-B-2Bb_0)/10$, 
$a_5=(2AC+2a_1b_2+2b_0a_3-a_3)/18$.
In this solution, the resonances $r=-1,0,3,6$ do not coincide one by 
one to the Kowalevski exponents $(-1,1,4,6)$ deduced by the Kowalevski 
matrix associated with the above balance. Nevertheless, the general 
solution (\ref{laubnt}), as well as any other solution work equally well 
in determining the matrix $A_f$. In our case, by performing the singular 
value decomposition of $A_f$, the subspace $ker(A_f)$ yields two independent 
integrals in involution, with coefficients (up to the computer precision) 
\begin{eqnarray}\label{intbnt}
\Phi_1 &= &0.211944988455(y^2 + p_y^2-4y^3) 
+ 0.049612730813(x^2+p_x^2) \nonumber \\
& &+ 0.216443010189(xp_xp_y-p_x^2y-x^2y^2 -{1\over 4}x^4) \\
& &-0.207446966722x^2y \nonumber
\end{eqnarray}
\begin{eqnarray}\label{intbnt2}
\Phi_2 &= &0.035960121414(y^2 + p_y^2-4y^3) + 0.342416478352(x^2+p_x^2) \nonumber \\
& &-0.408608475918(xp_xp_y-p_x^2y-x^2y^2 -{1\over 4}x^4) \\
& &-0.480528718746x^2y \nonumber
\end{eqnarray}
in terms of which we can express the Hamiltonian
\begin{equation}
H=2.1645645936295\Phi_1 + 1.146586289822\Phi_2
\end{equation}
and find a second integral orthogonal to the hamiltonian
\begin{equation}
I_2=-2.1645645936295\Phi_2 + 1.146586289822\Phi_1
\end{equation}
The integral $I_2$ is a linear combination of the hamiltonian and
of the integral $I_b$ given by Bountis et al. (1982)
\begin{equation}
I_2=0.1651752123636227(2H-{12\over 7}I_b)
\end{equation}

Let us note that an alternative way to obtain the matrix $A_f$ is 
by considering only the essential parts of the Laurent series 
(\ref{laubnt}), for as many different sets of values of the 
arbitrary parameters as requested in order to have a complete 
determination of the $M\times 38$ elements of $A_f$, with $M\geq 38$. 
This approach is preferable in computer implementations of the algorithm, 
because one does not need to calculate the terms of the Laurent series 
(\ref{laubnt}) beyond the highest positive resonance. Furthermore, 
Proposition 2, instead of Yoshida's theorem, can be used to select 
the quasiminomial basis set. 

A final remark concerns the applicability of Yoshida's condition  
$\nabla\Phi(b_i)\neq 0$. We checked whether this condition is satisfied 
in Hientarinta's table (1983) of integrable Hamiltonian systems of two 
degrees of freedom with a polynomial homogeneous potential. There are  
six non-trivial cases:\
$$
\mbox{(a)}~~~V(x,y)=2x^{3}+xy^{2}
$$
with integral $\Phi=yp_{x}p_{y}-xp_{y}^{2}+x^{2}x_{2}+\frac{1}{4}y^{4}$
$$
\mbox{(b)}~~~V(x,y)=\frac{16}{3}x^{3}+xy^{2}
$$
with integral  $\Phi=p_{y}^{4}+4xy^{2}p_{y}^{2}-\frac{4}{3}y^{3}
p_{x}p_{y}-\frac{4}{3}x^{2}y^{4}-\frac{2}{9}y^{6}$
$$
\mbox{(c)}~~~V(x,y)=2x^{3}+xy^{2}+i\frac{\sqrt{3}}{9}y^{3}
$$
with integral $\Phi=p_{y}^{4}+\frac{2}{\sqrt{3}}ip_{x}p_{y}^{3}+
\frac{2}{\sqrt{3}}iy^{3}p_{x}^{2}-(2y^{3}+2i\sqrt{3}xy^{2})p_{x}p_{y}+
(4i\sqrt{3}x^{2}y+2i\sqrt{3}y^{3}+4xy^{2})p_{y}^{2}+
\frac{4}{\sqrt{3}}ix^{3}y^{3}+\frac{2}{\sqrt{3}}ixy^{5}-
  x^{2}y^{4}-\frac{5}{9}y^{6}$
$$
\mbox{(d)}~~~V(x,y)=\frac{4}{3}x^{4}+x^{2}y^{2}+\frac{1}{12}y^{4}
$$
with integral $\Phi=yp_{x}p_{y}-xp_{y}^{2}+(\frac{1}{3}2x^{3}+xy^{2})y^{2}$
$$
\mbox{(e)}~~~V(x,y)=\frac{4}{3}x^{4}+x^{2}y^{2}+\frac{1}{6}y^{4}
$$
with integral $\Phi=p_{y}^{4}+\frac{2}{3}y^{4}p_{x}^{2}-
\frac{8}{3}xy^{3}p_{x}p_{y}+(4x^{2}y^{2}+\frac{2}{3}y^{4})p_{y}^{2}
+\frac{1}{9}(y^{8}+4x^{2}y{6}+4x^{4}y^{4})$
and
$$
\mbox{(f)}~~~V(x,y)=x^{5}+x^{3}y^{2}+\frac{3}{16}xy^{4}
$$
with integral $\Phi=yp_{x}p_{y}-xp_{y}^{2}+\frac{1}{2}x^{4}y^{2}+
\frac{3}{8}x^{2}y^{4}+\frac{1}{32}y^{6}$.

In all these cases, the principal balance, with all $b_i$ different from 
zero, satisfies Yoshida's condition. Note that case (f) the principal balance 
leads to `weak-Painlev\'{e}' solutions, but the associated integral $\Phi$ 
is easily recoverable by the present algorithm.  In fact, the role of 
Yoshida's condition is to exclude from Theorem 1 integrals which are 
composite functions of a simpler integral. For example, consider the integral 
$I=\Phi^2$, where $\Phi$ is the polynomial integral in any of the above six 
Hamiltonian systems. In all of them $\Phi$ is weighted-homogeneous for some 
weight $M$. Thus $I$ is weighted-homogeneous of weight $2M$. Substituting the 
special solution (\ref{balance}) in the integral $\Phi$ yields 
$\Phi(b_i/\tau^{\lambda_i}) = \tau^{-M}\Phi(b_i)$. Since $\Phi$ 
is an integral, it should be time-independent, thus $\Phi(b_i)=0$. 
Similarily, $I(b_i)=0$. However, while $\nabla\Phi(b_i)\neq 0$, we have 
$\nabla I=2\Phi\nabla\Phi$, thus $\nabla I(b_i)=0$. Thus, while $\Phi$ 
satisfies the Yoshida's condition, $I$ does not. This ensures that while 
$M$ is necessarily a Kowalevski exponent, the multiples of it, e.g. $2M$ 
are not necessarily Kowalevski exponents. Viewed under this context, it 
appears that the condition $\nabla\Phi(b_i)\neq 0$ makes a `natural' choice 
of the simplest integral among an infinity of possible integrals which are 
composite functions of $\Phi$. 

However, there are interesting counterexamples which challenge this point 
of view. One example is the homogeneous limit of the Holt (1982) 
Hamiltonian
\begin{equation}\label{hamholt}
H = {1\over 2}(p_x^2+p_y^2) -{3\over 4}x^{4/3}-y^2x^{-2/3}
\end{equation}
In this case, the potential is a homogeneous sum of quasi-monomials, of 
degree $4/3$. The second integral reads:
\begin{equation}\label{intholt}
\Phi=2\dot{y}^3 + 3\dot{y}\dot{x}^2 - 6\dot{y}y^2x^{-2/3} + 
9\dot{y}x^{4/3} - 18\dot{x}yx^{1/3}
\end{equation}
and it is weighted-homogeneous of degree $M=-6$. The principal 
balance is
\begin{equation}\label{bal1holt}
x=\left({1\over 3}\right)^{3/2}\tau^3,~~~
y=\left({i\over 3\sqrt{2}}\right)\tau^3 
\end{equation}
with resonances (=Kowalevski exponents) $r=-1,-2,-3,-4$. This case is 
remarkable because a) all the resonances are negative, and b) they 
are not equal to the weight of the integral $M=-6$. Thus, Yoshida's 
theorem does not apply in the case of this balance. Substituting the  
balance $b=((1/3)^{3/2},1/3\sqrt{2},3(1/3)^{3/2},1/\sqrt{2})$, i.e., 
Eq.(\ref{bal1holt}) in the gradient of the integral (\ref{intholt}), 
\begin{eqnarray}\label{gradphi}
\nabla\Phi(x,y,\dot{x},\dot{y}) &= 
&(4\dot{y}y^2x^{-5/3}+12\dot{y}x^{1/3}-6\dot{x}yx^{-2/3},\\ \nonumber
&~&-12\dot{y}yx^{-2/3}-18\dot{x}x^{1/3},6\dot{x}\dot{y}-18yx^{1/3},
6\dot{y}^2+3\dot{x}^2-6y^2x^{-2/3}+9x^{4/3})
\end{eqnarray}
yields $\nabla\Phi(b)=0$. Thus none of the resonances has to be equal 
to the weight of the integral. However, the form of $\Phi$ or $\nabla\Phi$ 
does not suggest that these functions are composite functions of some 
simpler integral. On the other hand, there is a second balance of the 
Hamiltonian $(\ref{hamholt})$, namely 
$b=((1/6)^{3/2},0,3(1/6)^{3/2},0)$, which corresponds to 
the dominant behavior
\begin{equation}\label{domholt}
x=\left({1\over 6}\right)^{3/2}\tau^3,~~~
y=A\tau^4 
\end{equation}
with $A$ arbitrary. The resonances here are $r=0,-1,-4$ and $-7$, 
but the Kowalevski exponents are $r_K = 1,-1,-4,-6$ (two of them 
differ by one from the respective resonances). Now, the Kowalevski 
exponent $r_k=-6$ is equal to the weight of the integral. If we 
look at $\nabla\Phi$, Eq.(\ref{gradphi}), we see that the components 
$\partial\Phi/\partial y$ and $\partial\Phi/\partial \dot{y}$ 
contain terms independent of $y$ and $\dot{y}$. Thus, for this 
particular balance $\nabla\Phi(b)\neq 0$, i.e., Yoshida's condition 
is satisfied. 

Thus, in the absence of a counterexample, we formulate the following\\
\\
{\bf Conjecture 3:} {\it In any scale-invariant system of the form 
(\ref{ode1},\ref{ode2}), which possesses a weighted-homogeneous 
first integral $\Phi$, at least one of the balances $b$ satisfies 
the condition $\nabla\Phi(b)\neq 0$.}\\
\\
Returning to the Holt Hamiltonian, the next step is the selection of 
quasi-monomial terms in the initial ansatz for a quasi-polynomial integral. 
Guided by the form of the Hamiltonian, natural exponents are adopted for 
the powers to which the momenta $p_x$, $p_y$, and of the variable $y$ are 
raised, while the variable $x$ is considered as raised to powers $m/3$, 
where $m$ is integer (positive or negative). Even under these restrictions, 
there is an infinity of possible quasi-monomial terms of weight -6. 
For example, the terms $p_ix^{m/3}y^{(2-m)/3}$, where $p_i$ 
is either $p_x$ or $p_y$, are of weight $-6$ for all $m\in Z$. Thus  
an arbitrary lower limit has to be set to $m$. This is chosen as the 
lowest bound of $m$ in the Hamiltonian, namely $m=-2$. Nevertheless, 
failure to find an integral with these restrictions on the quasi-monomial 
terms does not imply that an integral does not exist, because of the 
arbitrariness with respect to the lowest bound of negative exponents 
considered. This problem does not exist when there are only positive 
exponents present in the quasi-monomial terms of the equations of motion 
(or of the Hamiltonian). 

With the above restrictions, the quasi-monomial terms considered 
are:
$$
p_x^3,~p_x^2p_y,~p_xp_y^2,~p_y^3,~p_x^2x^{2/3},~p_x^2x^{-1/3}y
$$
$$
p_xp_yx^{2/3},~p_xp_yx^{-1/3}y,~p_y^2x^{2/3},~p_y^2x^{-1/3}y
$$
$$
p_xx^{4/3},~p_xx^{1/3}y,~p_xx^{-2/3}y^2,~p_yx^{4/3}
$$
$$
p_yx^{1/3}y,~p_yx^{-2/3}y^2,~x^2,x^y,y^2
$$

The final step is to build the matrix $A_f$ as in the previous 
examples, i.e. by replacing the Painlev\'{e} series in the initial 
ansatz for the integral. An interesting point is that in the case 
of the Holt Hamiltonian (\ref{hamholt}) we have to consider 
{\it left Painlev\'{e} series} (Pickering 1996), i.e., series 
in descending powers of $\tau$. This is because the balances 
$\sim \tau^3$ do not imply singular behavior as $\tau\rightarrow 0$. 
In this case, the limit $|\tau|\rightarrow\infty$ represents a 
singularity, but the left Painlev\'{e} series are convergent 
for all $\tau$ with $|\tau|>\epsilon$ for some real positive 
$\epsilon$. The series are constructed as:
\begin{equation}\label{leftser}
x_i(\tau)=\tau^{\lambda}\sum_{k=0}^{\infty}b_k\tau^{-k}
\end{equation}
where $\lambda>0$. Resonances and compatibilities are checked 
in the same way as in the usual Painlev\'{e} test. By this method, 
we were able to obtain the integral (\ref{intholt}) by a 
proper balancing of the quasi-monomial coefficients $c_i$ 
so as to eliminate the coefficients of the terms of successive 
descending powers of $\tau$ in the integral expression. 

As a final remark, it should be stressed that the selection of 
a quasi-polynomial ansatz for the integrals of a system of the 
form (\ref{ode1}), with the functions (\ref{ode2}) being 
quasi-polynomial, is not exhaustive. This can be easily 
exemplified in a case with polynomial functions. The 
Bogoyavlensky - Volterra B-type systems are given in normalized 
coordinates $u_i$ by the following set of autonomous nonlinear 
ODEs:
\begin{eqnarray}\label{odebog}
\dot{u_1} &= &u_1^2 +u_1u_2 \nonumber \\
\dot{u_i} &= &u_iu_{i+1}-u_iu_{i-1},~~~~~~~i=2,...,n-1 \\
\dot{u_n} &= &-u_nu_{n-1} \nonumber
\end{eqnarray}
the r.h.s. of Eqs.(\ref{odebog}) are homogeneous functions of second
degree in the variables $u_i$. The system (\ref{odebog})) admits 
balances of the form 
of the form:
\begin{equation}
u_i={a_i\over \tau}
\end{equation}
where $\tau = t-t_0$ is the time near a singularity
$t_0$ in the complex t-plane. In the case of the principal balance, 
the $a_i$ are non-zero solutions of the set of algebraic equations:
\begin{eqnarray}
-a_1 &= &a_1^2 +a_1a_2 \nonumber \\
-a_i &= &a_ia_{i+1}-a_ia_{i-1},~~~~~~~i=2,...,n-1 \\
-a_n &= &-a_na_{n-1} \nonumber
\end{eqnarray}
and they are given by the recursion formulas
\begin{equation}
a_{k+2}=a_k-1,~~~~~~a_1=(-1)^n[{n+1\over 2}],~~~~~~a_2=-1-a_1
\end{equation}
for $k=1,...,n$.

The resonances of the principal balance (=Kowalevski exponents) are 
given by the characteristic equation, i.e., setting the determinant 
of the Kowalevski matrix equal to zero. The determinant has a 
tridiagonal form, i.e., 
\begin{equation}
det\left(
\begin{array}{cccccccc}
a_1-r &a_1   &0   &0   &0 &0        &...   &0\\
-a_2  &-r    &a_2 &0   &0 &0        &      &0\\
0     &-a_3  &-r  &a_3 &0 &0        &      &0\\
.     &      &.   &.   &. &0        &      &.\\
.     &      &    &.   &. &.        &      &.\\
.     &      &    &    &. &.        &.     &.\\
0     &      &    &    &  &-a_{n-1} &-r    &a_{n-1}\\
0     &.     &.   &.   &  &0        &-a_n  &-r\\
\end{array}
\right)=0
\end{equation}
which can be solved easily yielding the resonances
\begin{equation}\label{resbog}
r_k=(-1)^kk,~~~~~~~~~k=1,...,n
\end{equation}

Assuming that the conditions for Yoshida's theorem hold, 
the weight of a weighted-homogeneous integral of (\ref{odebog}) 
should be one of the resonances (\ref{resbog}). Indeed, we find an 
integral by singularity analysis for {\it any} of the positive 
resonances given by equation (\ref{resbog}). The integrals are 
given by the recurrent relations:
\begin{equation}\label{intbog}
I_n^{(m)}=I_{n-1}^{(m)}+u_n^2I_{n-2}^{(m-2)} + 2\sum_{k=0}^{m/2-1}[
I_{n-m-1+2k}^{(2k)}\prod_{j=n-m+1+2k}^nu_j]
\end{equation}
where the convention $I_n^{(0)}=1$ and $I_n^{(m)}=0$ for 
all $m,n$ with $n=2,3,\ldots$ and $m>n$ is adopted.

For $n=3$, the resonances are $r=-3,-1,2$ and a polynomial integral is 
\begin{equation}\label{int2bog}
I_3^{(2)}=c^2=u_2^2 + u_3^2 + 2u_1u_2 + 2u_2u_3
\end{equation}
However, it is simple to see that this is not the only first integral 
of the system (\ref{odebog}). Defining $u=u_2+u_3$, and using any 
constant value $c$ of the integral (\ref{int2bog}), the equations 
of motion take the form
\begin{equation}\label{ode2bog}
\dot{u}={1\over 2}(u^2 - c^2)
\end{equation}
Integration of (\ref{ode2bog}) yields
\begin{equation}\label{solubog}
u=-c\coth({c(t-t_0)\over 2})
\end{equation}
Using $u$ instead of $u_2$ as a new independent variable yields the 
equation:
\begin{equation}\label{odeubog}
\dot{u}_3+uu_3=u_3^2
\end{equation}
Which can be solved for $u_3$ yielding 
\begin{equation}\label{solu3bog}
u_3=-c\frac{\cosh(c(t-t_0))-1}{\sinh(c(t-t_0))-c(t-t_0)+2c\gamma}
\end{equation}
where $\gamma$ is an integration constant. By eliminating the time 
between the solutions (\ref{solubog}) and (\ref{solu3bog}), a new 
first integral of the original equations is found: 
\begin{equation}\label{intlnbog}
I_{tr}=-2\gamma = {1\over c}\ln({u_2+u_3-c\over u_2+u_3+c}) 
+ \frac{2u_1 + u_2 + u_3}{u_1u_3}
\end{equation} 
which is a transcendental function of the variables $u_j$. This 
integral could not have been found by the initial polynomial 
ansatz. 

\section{Conclusions}

In this paper we have explored the question of whether it is possible 
to recover the first integrals of systems of first-order nonlinear 
ordinary different equations involving quasi-polynomial functions 
of the independent variables based on the information provided by 
singularity analysis. The main conclusions are:

a) The theorem of Yoshida (1983) constrains the choice of an 
initial ansatz for an integral with undetermined parameters,  
leaving, however, an infinity of possible choices. 

b) The condition of Yoshida's theorem ($\nabla\Phi(b_i)\neq 0$ and 
finite) holds for all the integrable Hamiltonian systems of two degrees 
of freedom included in Hietarinta's (1983) table, if $b_i$ is set equal 
to the principal balance $b_i\neq 0$ for all $i=1,\ldots 4$. Other 
types of balances have to be considered in more general systems.  

c) Substitution of the Painlev\'{e} series in a quasi-polynomial function 
$\Phi(x;c)$, where $c$ is the vector of undetermined parameters, allows 
to separate the terms in powers of the time and determine the parameters 
$c$ by singular value decomposition. Thus the information on 
quasi-polynomial integrals is contained in the Painlev\'{e}-type 
series solutions around any movable singularity.

d) In the case of balances $\tau^{\lambda}$ with $\lambda>0$, left 
Painlev\'{e} series must be used in the implementation of the algorithm.


\begin{thebibliography}{}
\bibitem{}
S. Abenda, V. Marinakis, T. Bountis.
On the connection between hyperelliptic separability and Painlev\'{e} 
integrability. 
J. Phys. A Math. Gen. 2001. 34. P. 3521-3539.  
\bibitem{}
M. Ablowitz, A. Ramani, H. Segur. 
A connection between nonlinear evolution equations and ordinary 
differential equations of P-type. I. 
J. Math. Phys. 1980. 21. P. 715-721.
\bibitem{}
V.I. Arnold.  Ed. Encyclopedia of Dynamical Systems Vol.III.  
Springer-Verlag. 1985. 
\bibitem{}
T. Bountis. H. Segur. F. Vivaldi. 
Integrable Hamiltonian systems and the Painlev\'{e} property.
Phys. Rev. A. 1982. 25. P. 1257-1264.
\bibitem{}
A.P. Fordy,  A. Pickering.
Analysing negative resonances in the Painlev\'{e} test.
Phys. Lett. A. 1991. 160. P. 347-354.
\bibitem{}
S.D. Furta.
On non-integrability of general systems of differential 
equations. 
Z. Angew. Math. Phys. 1996. 47. P. 112-131.
\bibitem{}
A. Goriely.  
Investigation of Painlev\'{e} property under time singularities 
transformations. J. Math. Phys. 1992. 33. P. 2728-2742.
\bibitem{}
A. Goriely. 
Integrability, partial integrability, and nonintegrability for 
systems of ordinary differential equations.
J. Math. Phys. 1996. 37. P. 1871-1893.
\bibitem{}
A. Goriely. Integrability and Non-integrability of Dynamical Systems. 
World Scientific. 2001. 
\bibitem{}
B. Grammaticos, B. Dorizzi, A. Ramani. 
Hamiltonians with high-order integrals and the 
``weak-Painlev\'{e}'' concept.
J. Math. Phys. 1984. 25. P. 3470-3473. 
\bibitem{}
G. Haller. Chaos near Resonance. Springer-Verlag. 1999. 
\bibitem{}
J. Hietarinta. 
A search for integrable two-dimensional hamiltonian systems with 
polynomial potential.
Phys. Lett. A. 1983. 96. P. 273-278.
\bibitem{}
J. Hietarinta.
Direct methods for the search of the second invariant.
Phys. Reports 1987. 147. P. 87-154.
\bibitem{}
C.R. Holt. 
Construction of new integrable Hamiltonians in two degrees of freedom
J. Math. Phys. 1982. 23. P. 1037-1046.
\bibitem{}
M. Lakshmanan, M. Senthil Velan. 
Direct integration of generalized Lie or dynamical symmetries of 
three degrees of freedom nonlinear Hamiltonian systems: Integrability 
and separability.
J. Math. Phys. 1992a. 33. P. 4068-4077.
\bibitem{}
M. Lakshmanan, M. Senthil Velan. 
Direct integration of generalized Lie symmetries of nonlinear 
Hamiltonian systems with two degrees of freedom: integrability and 
separability.
J. Phys. A: Math. Gen. 1992b.  25. P. 1259-1272.
\bibitem{}
J. Llibre, X. Zhang. 
Polynomial first integrals for quasi-homogeneous polynomial 
differential systems.
Nonlinearity 2002. 15. P. 1269-1280.
\bibitem{}
M. Marcelli, C. Nucci. 
Lie point symmetries and first integrals: the Kowalevsky top.
J. Math. Phys. 2003. 44. P. 2111-2132. 
\bibitem{}
K. Nakagawa. Ph.D Thesis. Tokyo Graduate University for 
Advanced Studies. 2002.
\bibitem{}
K. Nakagawa, H. Yoshida. 
A list of all integrable two-dimensional homogeneous polynomial 
potentials with a polynomial integral of order at most four in 
the momenta.
J. Phys. A: Math. Gen. 2001. 34. P. 8611-8630.
\bibitem{}
A. Pickering.
Weak perturbative Painlev\'{e} analysis and descending series 
solutions.
Phys. Lett. A. 1996. 221. P. 174-180.
\bibitem{}
A. Ramani, B. Dorizzi, B. Grammaticos. 
Painlev\'{e} Conjecture Revisited.
Phys. Rev. Lett. 1982. 49. P. 1539-1541.
\bibitem{}
A. Ramani, H. Yoshida, B. Grammaticos. 
Comment on 'Singular point analysis, resonances and Yoshida's theorem'.
J. Phys. A: Math. Gen. 1988. 21. P. 1471-1473.
\bibitem{}
D. Roekaerts, F. Schwarz. 
Painlev\'{e} analysis, Yoshida's theorems and direct methods 
in the search for integrable Hamiltonians.
J. Phys. A: Math. Gen. 1987. 20. P. L127-L133.
\bibitem{}
W.H. Steeb, J.A. Louw, M.F. Maritz. 
Singular point analysis, resonances and Yoshida's theorem.
J. Phys. A: Math. Gen. 1987. 20. P. 4027-4030.
\bibitem{}
A. Tsygvintsev. 
On the existence of polynomial first integrals of quadratic 
homogeneous systems of ordinary differential equations.
J. Phys. A: Math. Gen. 2001. 34. P. 2185-2193.
\bibitem{}
H. Yoshida. 
Necessary Condition for the Existence of Algebraic First Integrals 
- Part One - Kowalevski's Exponents.
Cel. Mech. 1983. 31. P. 363-379.
\bibitem{}
H. Yoshida. 
Necessary Condition for the Existence of Algebraic First Integrals
 - Part Two - Condition for Algebraic Integrability.
Cel. Mech. 1983. 31. P. 381-399.
\end{thebibliography}
\end{document}